\documentclass[12pt]{article}

\usepackage{amsmath}
\usepackage{graphicx}
\usepackage{amsfonts}
\usepackage{amssymb}
\usepackage{epsfig}
\usepackage{color}
\usepackage{cite}

\newcommand{\BE}{\begin{equation}}
\newcommand{\EE}{\end{equation}}
\newcommand{\om}{\omega}
\newcommand{\D}{\mathcal{D}}
\newcommand{\BEA}{\begin{eqnarray}}
\newcommand{\EEA}{\end{eqnarray}}
\newcommand{\F}{\Phi}
\newcommand{\FB}{\bar{\Phi}}
\newcommand{\lab}{\label}
\newcommand{\be}{\beta}
\newcommand{\te}{\theta}
\newcommand{\ch}{\cosh}
\newcommand{\sh}{\sinh}
\newcommand{\AB}{\bar{A}}
\newcommand{\ab}{\bar{a}}
\newcommand{\ka}{\kappa}
\newcommand{\HR}{\hat{R}}
\newcommand{\HT}{\hat{T}}
\newcommand{\RG}{\rangle}
\newcommand{\LG}{\langle}
\newcommand{\db}{\frac{d\be_1}{2\pi }}
\newcommand{\dbn}{\frac{d\be_n}{2\pi }}

\newcommand{\dt}{\frac{d\te_1}{2\pi }}
\newcommand{\dtn}{\frac{d\te_n}{2\pi }}
\newcommand{\mua}{\mu_{\alpha}}
\newcommand{\al}{\alpha}

\begin{document}
\setcounter{page}{0} \topmargin 0pt \oddsidemargin 5mm
\renewcommand{\thefootnote}{\arabic{footnote}}
\newpage
\setcounter{page}{0}

\begin{titlepage}

\begin{flushright}
SISSA REF 11/2003/CM
\end{flushright}

\vspace{0.5cm}
\begin{center}
{\large {\bf Massive ghost theories with a line of defects}}\\
\vspace{2cm}
{\large Paola Mosconi$^{1,2}$ } \\
\vspace{0.5cm} {\em $^{1}$International School for Advanced
Studies, Trieste, Italy }\\
\vspace{0.3cm} {\em $^{2}$Istituto Nazionale di Fisica Nucleare,
Sezione di Trieste} \\
\end{center}
\vspace{1cm}

\begin{abstract}
We study  free   massive fermionic ghosts, in the presence of an
extended line of impurities, relying on the Lagrangian formalism.
We propose two distinct defect interactions, respectively, of
relevant and marginal nature. The corresponding
scattering theories reveal the occurrence of resonances and
instabilities in the former case and the presence of poles with
imaginary residues in the latter. Correlation functions
of the thermal and disorder operators are computed exactly,
exploiting the bulk form factors and the matrix elements relative
to the defect operator. In the marginal situation, the one-point
function of the disorder operator displays a critical exponent
continuously varying with the interaction strength.
\end{abstract}

\end{titlepage}
\newpage

\section{Introduction}

After the seminal work by Ghoshal and Zamolodchikov
\cite{ghoshzam} on integrable field theories in the presence of a
boundary, a great deal of attention has been devoted to study
finite size effects, due especially to their numerous applications
to real physical problems. Quantum field theories with extended
line of defects\footnote{Actually, the critical behavior of
statistical systems with lines of defects has been widely studied
in the past years \cite{BE, IPT,bariev, mccoy, K, Brown, HP, BC,
KYP}. } generalize  these boundary models, introducing new and
original features \cite{defect,Leclair99,ludwlec,fring}.

The presence of impurities can be mimicked by the action of a \lq
defect\rq$\hspace{0.1cm}$ operator, placed along an infinite line
in the Euclidean space. In the continuum limit and away from
criticality, massive excitations can either participate to bulk
scattering processes or interact with the defect. In general, due
to the breaking of translational invariance, only reflection and
transmission are allowed. Such information can be encoded into a
scattering theory enriched by adding to the bulk S-matrix the
amplitudes relative to these two new processes. The integrability
of the model, originally studied in \cite{defect}, is guaranteed
by imposing the factorization condition which translates into a
set of cubic relations called the Reflection-Transmission
equations. In particular, it has been showed that, for diagonal
bulk scattering, non-trivial solutions for both the reflection and
transmission amplitudes can be found only in non interacting bulk
systems. In this light, free field theories play a prominent
r\^{o}le.

Recently, a wide interest has grown around free ghosts in two
dimensions, due to their relevance to the study of disordered
systems, polymer physics, quantum Hall states
\cite{cargese,marriage,sal92,salread,mooread} and above all as an
example of the simplest non-unitary/logarithmic conformal field
theories \cite{gurarie,flohr}. An exhaustive analysis of the
fermionic and bosonic ghosts' conformal field theories, possessing
respectively conformal charges $c=-2$ and $c=-1$, can be found in
\cite{fms86,kausch,betagamma}.

The main purpose of this work is to generalize a previously
studied model of free massive fermionic ghosts \cite{SFmass}, in
order to include the effects of inhomogeneities. In particular,
the knowledge of the scattering amplitudes (and the spectrum of
bulk excitations), along with general analyticity properties and
relativistic invariance, allows to reconstruct thoroughly the
off-shell dynamics, by computing exactly correlation functions.

The first step towards the realization of this program involves
the derivation of the transmission and reflection amplitudes. One
way to compute them consists in solving a bootstrap system of
equations (unitarity, crossing and factorization). However, in
this peculiar case, the absence of stringent constraints leaves a
broad arbitrariness in the choice of the solutions. Fortunately,
an alternative  description is possible, in terms of the
Lagrangian formalism
\BE\lab{action}\mathcal{A}=\mathcal{A}_B+g\int
d^2x\,\delta(x)\mathcal{L}_D(\varphi_i,\partial_y\varphi_i )\, ,
\EE where the bulk Euclidean action $\mathcal{A}_B$ and the
lagrangian density $\mathcal{L}_D$, encoding all the information
relative to the scattering processes on the defect line, both
depend on the local fields of the theory. According to the
strength of the coupling constant $g$, the line of inhomogeneities
can interpolate between a bulk and a surface statistical behavior.
If the defect interaction is relevant (irrelevant), a bulk
(surface) behavior is expected in the short distance limit, while
the marginal case shares both regimes. In the following,
a relevant and a marginal interactions are proposed and exact
expressions for the correlators of the most significant operators
in the theory are derived, by using the bulk form factors and the
matrix elements corresponding to the defect operator. In the
former case, resonance phenomena occur in the spectrum of
excitations, while the latter perturbation is responsible for
non-universal power-laws in the correlation functions of
operators, non-local in the ghost fields.

\section{Bootstrap approach}

The model we are going to study is that of free massive fermionic
ghosts \cite{SFmass} in the presence of an infinite line of
impurities placed at $x=0$, which, after a rotation in the
Minkowski plane, will be identified with the time axis.

The bulk spectrum of the theory is composed of a doublet of free
particles $A$ and $\AB$ with mass $m$, bearing respectively U(1)
charges $\pm1$.  Their scattering is ruled, in the bulk, by the
S-matrix $S=-1$. Due to the energy conservation, when a particle
hits the defect it can be either reflected or transmitted. All the
processes involved in the theory can be recast as a set of
algebraic equations \cite{defect}, relying on the algebra of the
Faddeev-Zamolodchikov operators. After the usual parameterization
of the particle's energy-momentum in terms of the rapidity
variable $(e,p)=(m\ch\te,m\sh\te)$, we associate to excitations of
type $\lq a$\rq $\hspace{0.1cm}$ the formal operator $A_a(\te)$
and to the defect line an operator $\D$, playing the r\^{o}le of a
zero rapidity particle, during the whole time evolution of the
system. The commutation relations, associated to the defect
algebra, read \BEA\lab{algebra}
A_a(\te)\D&=&R_a^b(\te)A_b(-\te)\D+
T_a^b(\te)\D A_b(\te) \, ,\nonumber\\
\D A_a(\te)&=&R_a^b(-\te)\D A_b(-\te)+T_a^b(-\te)A_b(\te)\D\,
,\EEA where, in the first equation, $R_a^b(\te)$ and $T_a^b(\te)$
denote, respectively, the reflection and transmission amplitudes
of an asymptotic particle $\lq a$\rq $\hspace{0.1cm}$ entering the
defect with rapidity $\te$, from the left. The second equation,
describing the scattering of a particle hitting the defect from
the right, is obtained from the first one, after an analytic
continuation $\te\to-\te$ in the rapidity variable. Consistency of
(\ref{algebra}) implies the unitarity conditions
\BEA\lab{unitarity}
R_a^b(\te)R_b^c(-\te)+T_a^b(\te)T_b^c(-\te)&=&\delta_a^c\, ,\nonumber\\
R_a^b(\te)T_b^c(-\te)+T_a^b(\te)R_b^c(-\te)&=&0\, .\EEA Crossing
relations read \BEA\lab{crossun}
&&\mathcal{C}^{aa''}R_{a''}^b\left(i\frac{\pi}{2}-\te\right)=S_{a'
b'}^{ab}(2\te)
\mathcal{C}^{b'b''}R_{b''}^{a'}\left(i\frac{\pi}{2}+\te\right)\, ,
\nonumber\\
&&T_a^b(\te)=\mathcal{C}^{bb'}T_{b'}^{a'}(i\pi-\te)\mathcal{C}_{a'a}\,
,\EEA with an antisymmetric charge conjugation matrix, such that
$\mathcal{C}^2=-1$.  As regards factorization conditions, the main
result of \cite{defect} guarantees that, for free theories
diagonal in the bulk, the Reflection-Transmission equations,
descending from integrability, are automatically satisfied.

At this point, solving the bootstrap system of equations
(\ref{algebra})-(\ref{crossun}), we are able in principle to
determine the scattering amplitudes $R_a^b$ and $T_a^b$. However,
a proliferation of solutions
occurs, due to the lack of constraints strong enough to fix the
reflection and transmission matrices in a closed form. A
simplified version of this model (i.e. a purely reflecting theory
which coincides with a boundary problem \cite{ghoshzam}) helps
visualizing the situation. Introduce the following
parameterization of the reflection matrix components:
\BEA\lab{param}
\begin{array}{cc}
  R_A^A(\te)=f(\te)R(\te) \hspace{0.5cm}& \hspace{0.5cm}R_A^{\AB}(\te)=g(\te)R(\te) \\
  R_{\AB}^{\AB}(\te)=f'(\te)R(\te) \hspace{0.5cm} &
  \hspace{0.5cm}R_{\AB}^{A}(\te)=g'(\te)R(\te)\, .
\end{array}\EEA
Consistency of the bootstrap system gives rise to the conditions
\BEA
&&R(\te)R(-\te)=[f(\te)f(-\te)+g(\te)g'(-\te)]^{-1}\nonumber\\
&&R(\te)R(-\te)=[f'(\te)f'(-\te)+g'(\te)g(-\te)]^{-1}\EEA \BEA
&&f(\te)g(-\te)+g(\te)f'(-\te)=0\nonumber\\
&&f'(\te)g'(-\te)+g'(\te)f(-\te)=0\EEA
\BE-\frac{g'\left(\frac{i\pi}{2}+\te\right)}{g'\left(\frac{i\pi}{2}-\te\right)}=
\frac{f'\left(\frac{i\pi}{2}+\te\right)}{f\left(\frac{i\pi}{2}-\te\right)}=
\frac{f\left(\frac{i\pi}{2}+\te\right)}{f'\left(\frac{i\pi}{2}-\te\right)}=
-\frac{g\left(\frac{i\pi}{2}+\te\right)}{g\left(\frac{i\pi}{2}-\te\right)}\,,\EE
which allow a richness of solutions. A comparison with the well
established theory of free massive Dirac fermions
\cite{ghoshzam,lecl95}, showing strong analogies with our ghost
system, is in order. Such model, obtained as a particular limit of
the Sine-Gordon one, admits non trivial boundary Yang-Baxter
equations, which provide a solution for the reflection amplitude
in terms of two parameters. In our case, starting directly from a
free theory, it is impossible to exploit factorization
constraints, in order to fix the form of the $R$-matrix.

\section{Lagrangian description}

To overcome the ambiguities, intrinsically concerned with the
bootstrap scenario, the lagrangian approach proves to be an
alternative route.

The Euclidean action, describing the bulk dynamics, is that of
free massless symplectic fermions \cite{kausch}, supplemented by a
mass term
\begin{equation}\label{actionbulk}
\mathcal{A}_B=\frac{1}{2} \, \int d^{2}x\, J_{\alpha \beta}\left(
\partial_{\mu}\Phi^{\alpha}
\partial^{\mu}\Phi^{\beta}+m^2\,
\Phi^{\alpha}\Phi^{\beta}\right)\, .
\end{equation}
$\Phi^\alpha$, which are zero dimensional anti-commuting fields
($\Phi$ and $\bar{\Phi}$), belong to the same doublet,
characterized by mass $m$, while $J_{\alpha\be}$ is an
antisymmetric tensor. A detailed analysis of the bulk system,
including mode expansions of the basic fields, commutation
relations and charge conjugation properties, can be found in the
Appendix A.

Inhomogeneities affect the bulk physics introducing a Lagrangian
density along the impurity line, according to (\ref{action}). A
relevant and a marginal interactions will be the object of our
study in order to derive explicit expressions for the reflection
and transmission amplitudes.

\subsection{Relevant perturbation}

Consider the system described by
\begin{equation}\label{Adm1}
\mathcal{A}=\mathcal{A}_B+ \frac{g}{2} \int\, d^2x\, \delta(x)\,
J_{\alpha\beta}\Phi^{\alpha}\Phi^{\beta}\, ,
\end{equation}
where the dimension of the coupling constant $g$ is $[mass]$. The
equations of motion read
\begin{eqnarray}\label{eqm1}
&&(\Box - m^2)\Phi=g \delta(x)\Phi \nonumber\\
&&(\Box - m^2)\bar{\Phi}=g \delta(x)\bar{\Phi}\, .
\end{eqnarray}
It is useful to split the fields into components belonging to the
two intervals $x<0$ and $x>0$ (after rotation to the Minkowski
space)
 \BEA\lab{dec}
\F(x,t)&=&\theta(x)\F_+(x,t)+\theta(-x)\F_-(x,t)\,  \nonumber\\
\FB(x,t)&=&\theta(x)\FB_+(x,t)+\theta(-x)\FB_-(x,t)\, ,\EEA in
order to derive the boundary conditions at $x=0$, given by
\begin{eqnarray}\label{bc1}
\Phi_+(0,t)-\Phi_-(0,t)&=&0 \, ; \nonumber\\
\partial_x(\Phi_+(0,t)-\Phi_-(0,t))&=&\,\frac{g}{4}(\Phi_+(0,t)+\Phi_-(0,t))
\end{eqnarray}
\BEA\lab{bc1bar} \FB_+(0,t)-\FB_-(0,t)&=&0\, ;\nonumber \\
\partial_x(\FB_+(0,t)-\FB_-(0,t))&=&\,\frac{g}{4}(\FB_+(0,t)+\FB_-(0,t))\,. \EEA
The mode expansions (\ref{modexp}), in terms of the operators $A$
and $\AB$ which interpolate the bulk excitations, allow us to
extract explicitly from (\ref{bc1})-(\ref{bc1bar}) the reflection
and transmission amplitudes
 \BEA\lab{Defmatr1} \left(\begin{array}{c}
  A_-(\be) \\
  \AB_-(\be) \\
  A_+(-\be) \\
  \AB_+(-\be)
\end{array}\right)=\left(\begin{array}{cc}
  R(\be,\ka) & T(\be,\ka) \\
  T(\be,\ka) & R(\be,\ka)
\end{array}\right)\left(\begin{array}{c}
  A_-(-\be) \\
  \AB_-(-\be) \\
  A_+(\be) \\
  \AB_+(\be)
\end{array}\right)\, ,
\EEA with
 \BEA\lab{RT1} R(\be,\ka)&=&\frac{1}{\sh\be+i\ka}\,\left(\begin{array}{cc}
  -i\ka & 0 \\
  0 & -i\ka
\end{array}\right) , \nonumber\\
T(\be,\ka)&=&\frac{1}{\sh\be+i\ka}\,\left(\begin{array}{cc}
  \sh\be & 0 \\
  0 & \sh\be
\end{array}\right)\, \EEA
and $\ka=g/4m$. $R$ and $T$, thus obtained, satisfy crossing and
unitarity conditions.

A strong analogy with the free bosonic theory, extensively treated
in \cite{defect}, emerges. A part from a doubling of the matrix
elements, the scattering amplitudes coincide. The main features
are the occurrence of resonances (i.e. unstable bound states
possessing a real part in the unphysical sheet, which do not
appear as asymptotic particles of the theory) for $\ka>1$ and
phenomena of instabilities for $\ka<-1$, characterized by poles
with imaginary part fixed at the value $i\pi/2$, acquiring an
increasing real part as $\ka$ is further depleted.

In the limit $g\to\infty$ ($\ka\to\infty$), corresponding to the
fixed boundary conditions $\Phi(0,t)=0$ and $\bar{\Phi}(0,t)=0$,
the defect line acts as a purely reflecting surface. On the
contrary, in the high-energy limit $\be\to\infty$, due to the
relevant character of the perturbation, the theory renormalizes to
a bulk regime, the impurity line becoming transparent.

\subsection{Marginal perturbation}

The Euclidean action
\begin{equation}\label{Adm0}
\mathcal{A}=\mathcal{A}_B-i g \int\, d^2x\,
\delta(x)\,(\Phi\partial_y\Phi+\bar{\Phi}\partial_y\bar{\Phi})\, ,
\end{equation}
where $g$ is a dimensionless coupling constant, describes the
effects of a marginal interaction on the defect line.
The equations of motion
\begin{eqnarray}\label{eqm0}
(\Box - m^2)\bar{\Phi}-2ig \delta(x)\partial\Phi=0 \\
(\Box - m^2)\Phi+2ig \delta(x)\partial\bar{\Phi}=0
\end{eqnarray}
lead to the following boundary conditions in the Minkowski plane
\BEA\lab{bc0bar}
\FB_+(0,t)-\FB_-(0,t)&=&0\, ; \nonumber \\
\partial_x(\FB_+(0,t)-\FB_-(0,t))&=& g\,\partial_t\F(0,t)\EEA
\BEA\lab{bc0} \F_+(0,t)-\F_-(0,t)&=&0\, ;\nonumber\\
\partial_x(\F_+(0,t)-\F_-(0,t))&=&-g\,\partial_t\FB(0,t)\, .\EEA
Exploiting again the mode expansions in terms of the operators $A$
and $\AB$, the reflection and transmission matrices assume the
form \BEA\lab{RT0} R(\be,\chi)&=&\frac{\sin\chi\ch\be}
{\ch^2\be-\cos^2\chi}\,
 \left(\begin{array}{cc}
  -\sin\chi\ch\be & -\cos\chi\sh\be \\
  \cos\chi\sh\be  & -\sin\chi\ch\be
\end{array}\right),\nonumber\\
T(\be,\chi)&=&\frac{\cos\chi\sh\be} {\ch^2\be-\cos^2\chi}\,
\left(\begin{array}{cc}
  \cos\chi\sh\be & -\sin\chi\ch\be \\
  \sin\chi\ch\be  & \cos\chi\sh\be
\end{array}\right)\, ,\
\EEA \BE \sin^2\chi=\frac{g^2}{4+g^2}\, .\nonumber\EE Some remarks
are in order. The action (\ref{Adm0}) is invariant under charge
conjugation, implemented by the transformations $\Phi\to
\bar{\Phi}$ and $\bar{\Phi}\to-\Phi$. Therefore, the relations
$R_A^A=R_{\AB}^{\AB}$ and $R_A^{\AB}=-R_{\AB}^{A}$, along with
their analogous counterparts for the transmission matrix, hold. On
the other hand, the U(1) symmetry, manifestly displayed by the
bulk action, is broken by the defect interaction. As a
consequence, scattering processes, which violate the conservation
of U(1) charges on the impurity line, can occur, allowing for
non-zero off-diagonal contributions. Exceptions to this behavior
concern the fixed ($g\to \infty $, $\cos\chi\to 0$) and the free
($g\to 0$, $\sin\chi\to 0$) boundary conditions, where a
restoration of the symmetry takes place.

Let us turn the attention on the analytic structure of the
reflection and transmission matrices. Since the theory is
non-unitary, a mechanism, akin to the one occurring in the scaling
Lee-Yang model \cite{YL}, is expected to take place. In other
words, residues, corresponding to poles in the scattering
amplitudes, are not supposed to be, a priori, real and positive.
This phenomenon is reminiscent of the non-hermitian nature of the
Hamiltonian associated to the system and does not contrast with
the unitarity requirement (\ref{unitarity}), preserving the
meaning of probability densities\footnote{Non-hermiticity of the
Hamiltonian implies, in particular, its left eigenstates $\LG
n_L|$ are not simply the adjoints of the right ones $|n_R\RG$.
Since, in addition, the Fock space states are also eigenstates of
the charge-conjugation operator with eigenvalues $(\pm i)^N$, $N$
being the particles' number, the relation $\LG n_L|=\LG n_R|\,
\mathcal{C}$ leads to the completeness condition
$\sum_n\,|n_R\RG\LG n_L|=\sum_n\,|n_R\RG\LG n_R|(\pm i)^n$. On the
other hand, eq. (\ref{unitarity}),
 relying only on the
assumption that in and out-kets, constructed in terms of the
asymptotic particles $A$ and $\AB$, form a basis in the Hilbert
space, is insensitive to hermiticity properties of the
Hamiltonian. }.

Poles appear both in the reflection and the transmission
amplitudes at $\be=i\chi$ and $\be=i(\pi-\chi)$, with
$\chi\in[0,\pi/2]$. In the case of diagonal matrix elements, the
corresponding residues give \BEA\lab{res}&& R_A^A\simeq
R_{\AB}^{\AB}\simeq T_{A}^{A}\simeq T_{\AB}^{\AB}\simeq
\frac{i}{2}\cdot
\frac{\sin\chi\,\cos\chi}{\be-i\chi} \nonumber\\
&&R_A^A\simeq R_{\AB}^{\AB}\simeq T_{A}^{A}\simeq
T_{\AB}^{\AB}\simeq\frac{i}{2}\cdot
\frac{-\sin\chi\,\cos\chi}{\be-i(\pi-\chi)}\, . \EEA Therefore,
the pole at $\be=i\chi$ is associated to a boundary bound state in
the direct channel, with positive binding energy $e_b=m\cos\chi$,
while the other one lives in the crossed channel. Since $e_b<m$
for every value of the coupling constant, the boundary bound
states are always stable and the theory is free of resonances and
instabilities of other nature. As regards off-diagonal processes,
the residues calculated  at $\be=i\chi$ assume the form \BEA
  &&R_{A}^{\AB}\simeq T_{A}^{\AB}\simeq \frac{i}{2}\cdot
\frac{i\,\sin\chi\,\cos\chi}{\be-i\chi} \hspace{1.2cm}
R_{\AB}^{A}\simeq T_{\AB}^{A}\simeq \frac{i}{2}\cdot
\frac{-i\,\sin\chi\,\cos\chi}{\be-i\chi}\; , \EEA while residues
computed in the crossed channel display an overall minus sign. As
mentioned before, the additional factor $\pm i$, appearing in the
numerator, is a consequence of the anomalous charge conjugation
properties of the ghost fields.

Finally, a comment on the marginal nature of the interaction:
performing the ultra-violet limit, except for peculiar values of
the coupling constant, all the scattering matrices' components
remain simultaneously finite.

\section{Correlation functions}

The problem at the heart of this paper concerns the computation of
correlation functions of the local fields $\phi_i(x,t)$, present
in the theory.

To realize this idea, in order to fully exploit the knowledge of
the bulk physics, it is worth performing a rotation in the
Minkowsi plane ($x\to -i t$, $t\to ix$), moving the defect line at
$t=0$.  In this new picture, the Hilbert space of states is the
same as in the bulk and the effects of impurities are taken into
account by an operator $\D$, placed at $t=0$, which acts on the
bulk states. Therefore,  correlation functions assume the form
\cite{defect} \BE\lab{corr}
\langle\Phi_1(x_1,t_1)...\Phi_n(x_n,t_n)\rangle =\frac{\langle 0|
T[\phi_1(x_1,t_1)...\mathcal{D}...\phi_n(x_n,t_n)]
|0\rangle}{\langle 0|\mathcal{D}|0\rangle}\, ,\EE
$\Phi_i(x_i,t_i)$ and $\phi_i(x_i,t_i)$ being the fields in the
Heisenberg representation, whose time evolutions are ruled,
respectively, by the exact Hamiltonian of the problem (bulk and
defect interactions) and the bulk Hamiltonian alone. As it appears
clearly, after inserting the completeness condition of the bulk
states in the right-hand side of (\ref{corr}), the above equation
can be computed only in terms of the Form Factors 
of the bulk fields and the matrix elements of the defect operator
on the asymptotic states. Another consequence of the axis-rotation
in the Minkowsi plane is the interchange of r\^{o}les between
energy and momentum. This affects the rapidity dependence of the
scattering amplitudes according to $\te\to (i\pi/2 -\te)$. In
compact notation it reads \BEA\lab{hat}
\HR^{ab}(\te)&=&\mathcal{C}^{aa'}R_{a'}^{b}\left(i\frac{\pi}{2}-\te\right)\,
, \nonumber\\
\HT^{ab}(\te)&=&\mathcal{C}^{aa'}T_{a'}^{b'}\left(i\frac{\pi}{2}-\te\right)
\mathcal{C}_{b'b}\,. \EEA

Let us recall here that asymptotic states are composed of neutral
pairs $A(\te)\AB(\be)$, obtained by acting with the corresponding
operators $A$ and $\AB$ on the vacuum $|0\RG$. Explicit
expressions for the bulk Form Factors have been derived in
\cite{SFmass}, while the simplest matrix elements of the defect
operator on the bulk states are \BEA\lab{matrixel2}&& \LG
A(\te)|\D|A(\te')\RG=2\pi\,\HT^{AA}(\te)\,\delta(\te-\te')\, ,
\nonumber\\
&&\LG \AB(\be)A(\te)|\D|0\RG=2\pi\,
\HR^{A\AB}(\te)\,\delta(\te+\be) \, , \nonumber\\
&&\LG 0|\D|A(\te)\AB(\be)\RG=-2\pi\, \HR^{A\AB}(\te-i\pi)\,
\delta(\be+\te-2\pi i)\, .\EEA In the remaining part of this
section, we are going to study correlators of the operator
\BE\lab{omdef}
\om(x,t)=\frac{J_{\alpha\be}}{2}\,\Phi^{\alpha}\Phi^{\be}(x,t) \,
, \EE associated to the massive perturbation of the critical bulk
theory, and the one-point function of the \lq disorder\rq
$\hspace{0.1cm}$ operator $\mu$.

\subsection {$\omega$ operator}

The simplest correlation function involving $\om$ is the one-point
function, defined as \BE\lab{om0}
\om_0(t,g)\equiv\LG\om(x,t)\RG=\sum_{n=0}^{\infty}\,\LG
0|\om(x,t)|n\RG\LG n|\D|0\RG\, ,\EE  the resolution of the
identity explicitly reading \begin{eqnarray}\label{resolution}
1=\sum_{n=0}^{\infty}\frac{1}{(n!)^{2}(2
\pi)^{2n}}\int_{-\infty}^{+\infty}d\theta_{1}...d\beta_{n}\,
|A(\theta_1),...,\AB(\beta_n)\rangle \langle
\AB(\beta_{n}),...,A(\theta_1)|\, .
\end{eqnarray}
Since $\om$ is the operator perturbing the critical theory in the
bulk, it turns out to be proportional to the trace of the
stress-energy tensor \cite{cardy88}. This implies, for free
theories, the remarkable property that only two-particle states
can be coupled to the vacuum \BE\lab{fftrace} \LG
0|\om(x,t)|A(\te_1)\AB(\be_1) \RG=2\pi\; e^{-m t
(\ch\be_1+\ch\te_1)+i m x (\sh \be_1+\sh \te_1)}\, .\EE Thus,
exploiting (\ref{matrixel2}), $\om_0$ can be recast as
\BE\lab{onepfom} \om_0(t,g)=2\int_0^\infty d\te
\,\HR^{A\AB}(\te)\, e^{-2mt\ch\te}\, .\EE Such formula is amenable
to discuss the different defect interactions.

For free boundary conditions, the reflection matrix is trivially
zero and the one-point function vanishes. In the case of fixed
boundary conditions, instead, $\HR^{A\AB}(\te)=-1$ and the short
distance limit is easily derived
\BE\lab{om0fix}\om_0(t)=-2\int_0^\infty d\te \, e^{-2mt\ch\te}=
-2K_0(2mt)\rightarrow 2\ln(mt) \, ,\hspace{0.8cm} mt\to 0 \,.\EE

Concerning the relevant perturbation, (\ref{onepfom}) assumes the
form \BE\lab{om0rel} \om_0(t,\kappa)=-\kappa
\int_{-\infty}^{+\infty}d\te \,
\frac{\exp[-2mt\ch\te]}{\ch\te+\kappa}\, .\EE In the limit of
fixed boundary conditions ($\ka\to \infty$) the previous result
(\ref{om0fix}) naturally follows while, in order to study the
large and short-distance regimes for arbitrary $\ka$, it could be
meaningful to manipulate a little bit the expression of $\om_0$.
The differential equation \BE\lab{diffeqrel}
\frac{\partial\om_0(t,\kappa)}{\partial(2mt)}-\kappa\om_0(2mt,\kappa)=2\kappa
K_0(2mt)\EE descending from (\ref{om0rel}), helps deducing the
large distance limit. Substituting the trial expansion
$\om_0(t,\ka)\sim e^{-2mt}(2mt)^{-\gamma}\sum a_\textit{l}\,
(2mt)^{-\textit{l}}$ into it, the asymptotic behavior $\om_0\sim
\frac{-2\ka}{1+\ka} \, K_0(2mt)$ is recovered as $mt\to \infty$.
On the other hand, the ultra-violet limit emerges more clearly if
we look at the expression \BE\om_0(t,\ka)=-2\ka\,
e^{(2mt)\ka}\int_{2mt}^{\infty}d\eta \,e^{-\eta
\ka}K_0(\eta)\,.\EE As far as $mt\to 0$, $\om_0$ always assumes
finite values. Summarizing, in the infra-red regime $\om_0$
follows the asymptotic behavior typical of the fixed boundary
conditions, while for small distances it remains finite,
approaching  zero as the coupling constant vanishes.

An analogous analysis can be performed for the marginal
interaction. The one-point function (\ref{onepfom}) specializes to
\BE\lab{om0mar}
\om_0(t,\chi)=-\sin^2\chi\int_{-\infty}^{+\infty}d\te \,
\frac{\sh^2\te}{\ch^2\te-\sin^2\chi}\;e^{-2mt\ch\te}\,.\EE The
corresponding differential equation \BE\lab{diffeqmar}
\frac{\partial^2\om_0(t,\chi)}{\partial(2mt)^2}-\sin^2\chi\,\om_0(t,\chi)=-2\sin^2\chi\,
\frac{K_1(2mt)}{2mt}\EE allows to derive both the asymptotic
limits. Exploiting a series expansion, as we did in the relevant
case, the low-energy regime leads to two different behaviors \BEA
\begin{cases}
    \om_0(t,\chi)\to e^{-(2mt)} (2mt)\sqrt{\frac{\pi}{2}}\;\frac{2 \sin^2\chi}{\sin^2\chi-1}\hspace{1cm}
    & \text{$\sin^2\chi\neq 1$}, \\
     \om_0(t,\chi)\to -2 K_0(2mt)& \text{$\sin^2\chi=1$}.
  \end{cases}
\EEA As regards the ultra-violet limit, $\om_0(t,\chi)\sim
2\sin^2\chi\ln(2mt)$.

We turn now the attention to the two-point functions involving the
operator $\om$.  Two different situations can occur.

\vspace{0.6cm}

Consider the case in which the operators lie on opposite sides of
the defect line, i.e. $t_1<0$ and $t_2>0$. The correlator is given
by \BE \lab{G1} G_1(\rho_1,\rho_2;g)=\sum_{i,j}\, \LG
0|\om(\rho_2)|i\RG \LG i|\D|j\RG \LG j|\om(\rho_1)|0\RG \, ,\EE
with the collective variable $\rho=(x,t)$.  As before, the series
contains only a finite number of terms. In order to perform the
calculations, we need the expression of the \lq defect\rq
$\hspace{0.1cm}$ matrix element involving four particles \BEA
\lab{me2per2}
\LG\AB(\be_1)A(\te_1)|\D|A(\te'_1)\AB(\be'_1)\RG&=&(2\pi)^2\;
[\HT^{AA}(\te_1)\HT^{\AB\AB}(\be_1)\,
\delta(\te_1-\te'_1)\delta(\be_1-\be'_1)+\nonumber\\
&-&\HR^{A\AB}(\te_1)
\HR^{A\AB}(\te'_1-i\pi)\,\delta(\be_1+\te_1)\delta(\be'_1+\te'_1-2\pi
i )]\, . \EEA Introducing a redefinition of variables in terms of
$t\equiv t_2-t_1$ and $x\equiv x_2-x_1$, we finally obtain
\BE\lab{G1rel} G_1(\rho_1,\rho_2;\kappa)=-\left[ \frac{\partial
F(mx,mt;\kappa)}{\partial(mt)}\right]^2+\om_0(t_1,\kappa)\om_0(t_2,\kappa)\,
, \EE
 \BE\lab{Frel}F(x,t)=\int_{-\infty}^{+\infty}d\te\,
\frac{\exp[-t\ch \te+ix\sh\te]}{\ch\te+\kappa}\, \EE for the
relevant perturbation and
\BE\lab{G_1mar}G_1(\rho_1,\rho_2;\chi)=-\cos^4\chi\,\left[
\frac{\partial^2F(mx,mt;\chi)}{\partial(mt)^2}\right]^2+
\om_0(t_1,\chi)\om_0(t_2,\chi)\,, \EE
\BE\lab{Fmar}F(x,t)=\int_{-\infty}^{+\infty}d\te\,
\frac{\exp[-t\ch \te+ix\sh\te]}{\ch^2\te-\sin^2\chi}\,, \EE for
the marginal one. In the limit of an infinitely reflecting surface
($\ka\to \infty$ and $\cos^2\chi\to 0$), only the vacuum
expectation values of the two $\om$ operators survive.

\vspace{0.6cm}

Another situation can happen, in which the two $\om$ operators
reside on the same half of the Minkowsi plane. Let us consider,
for convenience, $t_2\geq t_1>0$ and define $t\equiv t_2-t_1$,
$\bar{t}\equiv t_2+t_1$, $x\equiv x_2-x_1$,
$r\equiv\sqrt{x^2+t^2}$. The general expression for the two-point
function is \BE\lab{G2}G_2(\rho_1,\rho_2;g)= \sum_{i,j} \, \LG
0|\om(\rho_2)|i\RG\LG i|\om(\rho_1)|j\RG\LG j|\D|0\RG\, .\EE
Following the lines traced in \cite{defect}, after straightforward
calculations, we end up with
\BE\lab{G2rel}G_2(\rho_1,\rho_2;\kappa)=-[2K_0(mr)+\kappa F(m\bar
t,mx)]^2+\om_0(t_1,\kappa)\om_0(t_2,\kappa)\, ,\EE in the relevant
case and \BE\lab{G2mar} G_2(\rho_1,\rho_2;\chi)=-\left[
2K_0(mr)+\sin^2\chi\,\frac{\partial^2F(m\bar
t,mx)}{\partial(mx)^2}\right]^2+\om_0(t_1,\chi)\om_0(t_2,\chi)\, ,
\EE for the marginal perturbation. As it appears clearly, the
solutions found are invariant under translations along the
$x$-axis, consistently with the picture adopted, which preserves
momentum.

\subsection{Disorder operator}

Finally, we examine the one-point function of the operator $\mu$,
which is only a specific example belonging to the widest class of
the \lq disorder\rq $\hspace{0.1cm}$operators, non-local with
respect to the ghost fields. The analysis concerning the leading
behavior of their correlators, which relies on a \lq cluster\rq
$\hspace{0.1cm}$expansion, is the main purpose of the Appendix C,
while  a detailed discussion about them in bulk free theories can
be found in \cite{kyoto,ber97,grinza} (ordinary complex fermions
and bosons) and \cite{SFmass} (fermionic and bosonic ghost
systems). The one-point correlator can be written as follows
\BE\lab{mugen} \mu_0(t,g)\equiv\LG
\mu(x,t)\RG=\sum_n\,\LG0|\mu(x,t)|n\RG\LG n|\D|0\RG\,.\EE In this
case, $\mu$ couples the vacuum to neutral states, composed of an
even number of excitations. As a consequence, the sum does not
truncate and, to explicitly evaluate (\ref{mugen}), matrix
elements involving an arbitrary (even) number of particles
\BEA\lab{me2n}
\LG\AB(\be_n)...\AB(\be_1),A(\te_n)...A(\te_1)|\D|0\RG=
(-)^\frac{n(n-1)}{2}\,
n!\,(2\pi)^n\prod_{k=0}^{n}\HR^{A\AB}(\be_k)\, \delta(\be_k+\te_k)
\EEA are required. In addition, since the defect operator $\D$ is
responsible for processes involving only absorption or emission of
couples of particles with opposite rapidities,
 $\mu_0$ finally assumes the form
\BEA\lab{mu0}\mu_0(t,g)=\sum_{n=0}^\infty\;\frac{(-)^{n(n-1)/2}}{n!}&&\int
 \db\,...\,\dbn \prod_{k=1}^{n}\left[
\HR^{A\AB}(\be_k)e^{-2mt\ch\be_k}\right] \cdot\nonumber\\
&&\cdot\, f_n^{1/2}(-\be_1,...,-\be_n,\be_1,...,\be_n)\, .\EEA
Exact expressions for the bulk Form Factors are given in
\cite{SFmass} \BE\lab{ff}
f_n^{1/2}(\te_1,...,\te_n,\be_1,...,\be_n)=\LG
0|\mu_{1/2}(0)|A(\te_1)...A(\te_n)\AB(\be_1)...\AB(\be_n)\RG=(-)^{n(n+1)/2}
|A_n|\, , \EE where $|A_n|$ denotes the determinant of a matrix
whose components read \BE
A_{ij}=\frac{1}{\ch\frac{\te_i-\be_j}{2}}\nonumber\;.\EE In order
to discuss the  effects due to the different interactions
localized along the defect line, (\ref{mu0}) proves to be a good
starting-point.

Again, free boundary conditions lead to the trivial solution
$\mu_0=0$.

In the case of fixed boundary conditions, it is possible to
recover the leading short-distance behavior of the one-point
function, in an exact way. The details of the calculation will be
postponed to the Appendix B, while here only the main results will
be given. Since the reflection matrix component $\HR^{A\AB}$ is
trivially $-1$, exploiting the theory of Fredholm determinants
\cite{schwinger}, $\mu_0$ can be recast as
\BE\lab{fredfix}\mu_0(t)=\sum_{n=0}^{\infty}\frac{1}{n!}
\int_{-\infty}^{+\infty}\dt...\dtn\;
 e^{-2mt\sum_1^n\,\ch\te_k}\;
|A_n|=\det\left(1+\frac{1}{\pi}\,V(t)\right)\,,\EE where the
kernel is given explicitly by \BEA\lab{kerfix}&&V(\te_i,\te_j,t)=
\frac{e(\te_i,t)\,e(\te_j,t)}{2\ch\frac{\te_i+\te_j}{2}}\; ,\nonumber\\
&&e(\te,t)= e^{-mt\ch\te}\,.\EEA Alternatively, $\mu_0$ can be
expressed as \BE\lab{det}\mu_0(t)=\prod_{i=1}^{\infty}\,
\left(1+\frac{1}{\pi}\, \lambda^{(i)}(t)\right)^{a_i(t)}\, ,\EE
where $\lambda_i$ are the eigenvalues of the integral operator
$V(t)$, distributed with multiplicity $a_i(t)$. As far as $mt$ is
finite, $V(t)$ is a square integrable operator possessing a
discrete spectrum. However, in the short-distance limit, $mt\to
0$, this condition ceases to hold and the eigenvalues become dense
in the interval $(-\infty,+\infty)$, with a multiplicity growing
logarithmically as $\sim \ln\frac{1}{mt}$. Therefore, the disorder
operator $\mu$ follows the leading power-low behavior
\BE\lab{uvexp}\mu_0(t)\sim\frac{C}{(2t)^{x_\mu}}\,,\EE with
$x_{\mu}=-1/4$. This result is consistent with the intuitive idea
that, upon approaching the impurity line in the ultra-violet
limit, the operator $\mu$, characterized by the conformal weight
$(-\frac{1}{8},-\frac{1}{8})$, starts interacting with its mirror
image on the other side of the defect, along the identity channel.
As a final remark, we hint at the possibility of sub-leading
logarithmic corrections.

As regards the effects produced by the relevant perturbation,
(\ref{mu0}) behaves as
\BEA\lab{murel}\mu_0(t;\kappa)=\sum_{n=0}^{\infty}\frac{1}{n!}\left(
\frac{1}{\pi}\right)^n\int_{-\infty}^{+\infty}d\te_1...d\te_n\left[\prod_{k=1}^n\,\frac{\kappa\,
e^{-2mt\ch\te_k} }{2\,(\ch\te_k+\kappa)}\right]
|A_n|=\det\left(1+\frac{1}{\pi}V(t;\kappa)\right) , \EEA with the
kernel \BEA\lab{ker1}&&V(\te_i,\te_j,t;\kappa)=
\frac{e(\te_i,t;\kappa)\,e(\te_j,t;\kappa)}{2\ch\frac{\te_i+\te_j}{2}}\; ,\nonumber\\
&&e(\te,t;\kappa)=\sqrt{\frac{\kappa}{\ch\te+\kappa}}\cdot
e^{-mt\ch\te}\,.\EEA In the short-distance limit,  $|V|^2$ becomes
unbounded, the leading singularity being dictated by the fixed
boundary conditions' one. Thus we find the same critical exponent
as in the previous case.

More interesting is the marginal situation. From general
considerations extrapolated from the Ising model
\cite{bariev,mccoy}, the non-universal nature of the marginal
interaction is expected to affect the non-local sector of the
theory, inducing a critical exponent continuously dependent on the
coupling constant. Indeed, $\mu_0$ assumes the form
\BEA\lab{mumar}\mu_0(t;\chi)&=&\sum_{n=0}^{\infty}\frac{1}{n!}\left(
\frac{\sin^2\chi}{\pi}\right)^n\int_{-\infty}^{+\infty}d\te_1...d\te_n\left[\prod_{k=1}^n\,\frac{\sh^2\te_k\;
e^{-2mt\ch\te_k} }{2\,(\ch^2\te_k-\sin^2\chi)}\right]
|A_n|=\nonumber\\
&=&\det\left(1+\frac{\sin^2\chi}{\pi}\,V(t;\chi)\right)\,,\EEA
where\BEA\lab{ker0}&&V(\te_i,\te_j,t;\chi)=
\frac{e(\te_i,t;\chi)\,e(\te_j,t;\chi)}{2\ch\frac{\te_i+\te_j}{2}}\; ,\nonumber\\
&&e(\te,t;\chi)=\sqrt{\frac{\ch^2\te-1}{\ch^2\te-\sin^2\chi}}\cdot
e^{-mt\ch\te}\,.\EEA Repeating an analysis similar to the one
carried out for the fixed boundary condition, but this time with a
 parameter depending on the coupling constant, in front of the kernel in
(\ref{mumar}), we finally obtain the critical exponent
\BE\lab{expmar}x_{\mu}=\frac{1}{4}-\frac{1}{2\pi^2}\;[\arccos^2(\sin^2\chi)+\arccos^2(-\sin^2\chi)]\,.\EE

\section{Final remarks}

In this paper we have studied the effects induced by a defect
interaction on the free theory of massive fermionic ghosts.

Working in the Lagrangian approach, we have dealt with two defect
perturbations, respectively, of relevant and marginal nature.
Explicit expressions for the reflection and transmission matrices
have been derived. A careful analysis of their excitation spectra
has pointed out the possibility of resonances and instabilities in
the former case, and the occurrence of imaginary residues,
relative to poles in the scattering amplitudes, in the latter one.
Successively, we turned our attention to the exact computation of
correlation functions, involving the most interesting operators in
the theory, i.e. $\om$, local in the ghost fields, and
 $\mu$, belonging to one of the
non-trivial sectors of the model. In the marginal situation, a
non-universal behavior in the one-point function of the \lq
disorder\rq $\hspace{0.1cm}$ operator $\mu$ has clearly emerged.
Finally, the last appendix has been devoted to the analysis of the
most general \lq disorder\rq $\hspace{0.1cm}$fields $\mu_\al$,
characterized by non-locality index $\al$. The leading
short-distance behavior of their one-point function has been
investigated by means of the \lq cluster\rq
$\hspace{0.1cm}$expansion \cite{smir89,cluster}.

It is worth noticing that a delicate point of the present
discussion concerns the comparison between the bootstrap approach
and the Lagrangian description, in order to derive explicit
expressions for the reflection and transmission amplitudes. In the
former case, a richness of solutions descends but their physical
explanation and \lq classification\rq, in terms of a fixed number
of parameters related to the bulk S-matrix, results problematic.
On the other hand, the  Lagrangian approach, though subjected to
the strong restriction of dealing  only with local interactions,
allows for a limited number of solutions,  amenable of an easiest
control. For instance, besides the defect perturbations already
introduced, analyzing other kind of interactions could help
identifying new boundary conditions and, possibly, the operator
content of the boundary theory.

Finally, we conclude with a remark on the simplified problem of a
pure reflecting surface. As hinted at the end of the second
section in relation to the free Dirac massive fermions, free
theories, derived as limit of interacting ones, admit a richer
structure, as it appears clearly in the bootstrap approach. It
would be tempting, in this boundary case, to find an interacting
theory, if any, behind the fermionic ghost model.

\begin{flushleft}\large
\textbf{Acknowledgements}
\end{flushleft}
I would like to thank G. Mussardo for suggesting me the subject of
this work and for many valuable discussions.
This work was supported
by the TMR Network \lq\lq EUCLID. Integrable models and applications:
from strings to condensed matter\rq\rq, contract number HPRN-CT-2002-00325.

\section*{Appendix A}

In this section,  some useful results on the bulk system of
fermionic ghosts are collected. The action is described by Eq.
(\ref{actionbulk}) where the symplectic form $J_{\alpha\beta}$
reads explicitly
\begin{equation}\label{J}
J_{-+}=-J_{+-}=1  \hspace{2mm}, \hspace{2mm}
J_{\alpha\gamma}J^{\gamma\beta}=\delta^{\beta}_{\alpha}\,,
\end{equation}
and the ghost fields $\Phi^{\pm}$, for later convenience, are
redefined according to
\begin{eqnarray}
&&\Phi^{+}\to \Phi \nonumber\\
&&\Phi^{-}\to \bar{\Phi}\,. \nonumber
\end{eqnarray}
The mode expansions for the components $\Phi_{(\pm)}$ and
$\bar{\Phi}_{(\pm)}$, previously introduced (\ref{dec}), are
\BEA\lab{modexp} \F_{(\pm)}(x,t)&=&\int d\be
\left[\ab_{(\pm)}(\be)
e^{-im(t\ch\be-x\sh\be)}+a_{(\pm)}^\dag(\be)e^{im(t\ch\be-x\sh\be)}\right]\,
, \nonumber \\
\FB_{(\pm)}(x,t)&=&\int d\be \left[-a_{(\pm)}(\be) e^{-im
(t\ch\be-x\sh\be)}+\ab_{(\pm)}^\dag(\be)e^{im(t\ch\be-x\sh\be)}\right]\,,
\EEA where the creation and annihilation operators are subjected
to the anti-commutation relations \BEA\lab{anticomm}
&&\{a_{(\pm)}(\be),a^{\dag}_{(\pm)}(\be\rq)
\}=2\pi\delta(\be-\be\rq)\hspace{1.6mm},\hspace{1.6mm}\{a_{(\pm)}(\be),a_{(\pm)}(\be\rq)
\}=0=\{a_{(\pm)}^{\dag}(\be),a^{\dag}_{(\pm)}(\be\rq) \}
;\nonumber\\
&&\{\ab_{(\pm)}(\be),\ab^{\dag}_{(\pm)}(\be\rq)\}=2\pi\delta(\be-\be\rq)
\hspace{1.6mm},\hspace{1.6mm}\{\ab_{(\pm)}(\be),\ab_{(\pm)}(\be\rq)
\}=0=\{\ab_{(\pm)}^{\dag}(\be),\ab^{\dag}_{(\pm)}(\be\rq) \}\,.
 \EEA
Charge conjugation implemented on the Fock operators
\BEA\lab{conj}
\mathcal{C}a(\be)\mathcal{C}^{-1}&=&\ab(\be)\hspace{4mm},\hspace{4mm}
\mathcal{C}a^{\dag}(\be)\mathcal{C}^{-1}=\ab^{\dag}(\be)\,
;\nonumber\\
\mathcal{C}\ab(\be)\mathcal{C}^{-1}&=&-a(\be)\hspace{2mm},\hspace{2mm}
\mathcal{C}\ab^{\dag}(\be)\mathcal{C}^{-1}=-a^{\dag}(\be)\,,\EEA
induces the following transformations on the ghost fields
$\Phi\to\bar{\Phi}$ and $\bar{\Phi}\to-\Phi$. Finally, it is
useful, for  notational reasons, to identify the operator creating
the bulk excitations with the excitations themselves
\BEA\lab{masspart} &&a^{\dag}(\be)\to A(\be)\, , \nonumber\\
&&\ab^{\dag}(\be)\to \AB(\be)\,.\EEA

\section*{Appendix B}

In this appendix we evaluate the critical exponent of the disorder
operator $\mu$, corresponding to the fixed boundary conditions.
Let us consider the logarithm of Eq. (\ref{det}) \BE\lab{ln} \ln
\mu_0(t)=\sum_{i=1}^{\infty}\,a_i(t) \ln\left(1+\frac{1}{\pi}\,
\lambda^{(i)}(t)\right)\,, \EE where, as explained before,
$\lambda_i(t)$ are the eigenvalues of the integral operator
$V(t)$, defined by Eq. (\ref{kerfix}). In the limit $mt\to 0$,
such operator turns out to be singular (it loses the property of
square-integrability) and consequently, its eigenvalues become
dense in $(-\infty,+\infty)$. The first problem to be faced
concerns finding the exact solution to the eigenvalue equation
 \BE\lab{inteq}
\int_{-\infty}^{+\infty}d\te_2\,\frac{1}{2\ch\frac{\te_1+\te_2}{2}}\,\phi(\te_2)=\lambda\,\phi(\te_1)\,,\EE
which, after proper changes of variables, assumes definitely the
form \BE \lab{uv}\int_0^\infty du
\,\frac{1}{uv+1}\,\xi(u)=\lambda\,\xi(v)\,.\EE The peculiar
expression of the new kernel $K(u,v)=\frac{1}{uv+1}$ suggests to
consider the Mellin transform of both sides of (\ref{uv})
\cite{makarenko,ditkin}. We finally end up with a simpler
eigenvalue equation for the transformed quantities
\BEA(\lambda^2-\tilde{K}(s)\tilde{K}(1-s))\,\tilde{\xi}(s)=0\, ,
\EEA where
\BE \tilde{K}(s)=\frac{\pi}{\sin\pi s}\hspace{0.4cm},
\hspace{0.4cm}0<\Re e\,s<1\,.\EE Some comments could be useful to
evaluate the spectrum. Since the kernel is a symmetric function of
its arguments and it is bounded, the spectrum has to be real and
limited. Hence \BE\lab{eigenvalues}
\lambda_{\pm}(\tau)=\frac{\pm\,\pi}{\ch\pi\tau}\hspace{0.4cm},
\hspace{0.4cm}\tau\in (-\infty,+\infty)\,.\EE Now, Eq. (\ref{ln})
assumes the form \BE \lab{log}
\ln\mu_0(t)=a(t)\int_{-\infty}^\infty d\tau
\left[\ln\left(1+\frac{1}{\pi}\,\lambda_+(\tau)\right)+
\ln\left(1+\frac{1}{\pi}\,\lambda_-(\tau)\right)\right]\,, \EE
where the multiplicity has been assumed to be uniform. Moreover,
thanks to Mercer's theorem, $a(t)\sim
\frac{1}{2\pi}\ln\frac{1}{t}\hspace{0.1cm}$.  In the end, the
critical exponent is given by \cite{gra}
\BEA\lab{exp}x_{\mu}&=&\frac{1}{2\pi}\int_{-\infty}^\infty d\tau
\left[\ln\left(1+\frac{1}{\ch\pi\tau}\right)+\ln\left(1-\frac{1}{\ch\pi\tau}\right)\right]=\nonumber\\
&=&\frac{1}{\pi^2}\left[\frac{\pi^2}{4}-\frac{1}{2}\,(\arccos^2(1)+\arccos^2(-1))\right]=-\frac{1}{4}\,.\EEA

\section*{Appendix C}

In this last appendix we discuss  generic \lq disorder\rq
$\hspace{0.1cm}$operators $\mu_\alpha$, which pick up a
non-locality phase $e^{\pm 2\pi i\alpha}$, when they are taken
around the ghost fields in the Euclidean plane \BEA \Phi(z\,
e^{2\pi i},\bar{z}\, e^{-2\pi i})\mu_\alpha(0,0)&=&e^{2\pi i
\alpha}\Phi(z,\bar{z})\mu_\alpha(0,0)\,,\nonumber\\
\bar{\Phi}(z\, e^{2\pi i},\bar{z}\, e^{-2\pi
i})\mu_\alpha(0,0)&=&e^{-2\pi i
\alpha}\bar{\Phi}(z,\bar{z})\mu_\alpha(0,0)\, .\EEA In particular,
we are interested in deriving the leading short-distance behavior
of their one-point function in the case of fixed boundary
conditions, in order to perform a comparison with the exact result
previously obtained for the specific value $\alpha=\frac{1}{2}\,$.

The starting point is Eq. (\ref{mu0}), where the Form Factors
$f_n^{1/2}(-\be_1,...,\be_n)$ must be replaced by the expression
\cite{SFmass} \BE
f_n^{\alpha}(-\be_1,...-\be_n,\be_1,...,\be_n)=(-)^{n(n+1)/2}\,(\sin
\pi\alpha)^n\;e^{-\left(\alpha-\frac{1}{2}\right)\sum_i^n \,
2\be_i}\, |A_n|\,,\EE with $|A_n|$ the determinant of the $n\times
n$ matrix, whose components satisfy \BE
A_{ij}=\frac{1}{\ch\frac{\be_i+\be_j}{2}}\,.\EE In a compact form,
we can rewrite \BE\lab{mua0}\mu_{0}^{\al}(t)\equiv\LG\mua
(x,t)\RG=\sum_{n=0}^{\infty}\,\frac{1}{n!}\,\int_{-\infty}^{+\infty}\,
d\be_1...d\be_n \,
e^{-\rho\sum_j^n\ch\be_j}\,g_n^\alpha(\be_1,...,\be_n)\,,\EE where
$\rho=2mt$ and \BE\lab{ga}g_n^\alpha(\be_1,...,\be_n)\equiv
\frac{1}{(2\pi)^n}\,(\sin\pi\alpha)^n\,e^{-\left(\al-\frac{1}{2}\right)\sum_{j}^{n}2\be_j}\,|A_n|\,.\EE
These last two relations appear  suitable to perform a \lq
cluster\rq $\hspace{0.1cm}$expansion, according to the technique
exposed, for instance, in \cite{smir89,cluster}. Therefore, the
logarithm of (\ref{mua0}) assumes the form \BE\lab{logmua} \ln
\mu_0^{\al}(t)=\sum_{n=1}^{\infty}\, \frac{1}{n!}\,
\int_{-\infty}^{+\infty}\,d\be_1...d\be_n\,
e^{-\rho\sum_j^n\ch\be_j}\,h_n^{\al}(\be_1,...,\be_n)\,,\EE where
the functions $h_n^{\al}$ are proper combinations of the
$g_n^\alpha$. For our purposes, we need only the first few
relations, which read explicitly \cite{cluster}
\BEA\lab{gf}
g_1^\al&=&h_1^\al\nonumber\\
g_{12}^\al&=&h_{12}^\al+h_1^\al h_2^\al\nonumber\\
g_{123}^\al&=&h_{123}^\al+ h_{12}^\al h_3^{\al}+ h_{23}^\al
h_1^{\al}+h_{31}^\al h_2^{\al}+ h_1^\al h_2^\al h_3^\al\,.\EEA

The key point of the standard \lq cluster\rq
$\hspace{0.1cm}$expansion is that, since the functions $h_n$
depend only on rapidity differences, they contain a redundant
variable. Thus, it is possible, at all orders, to extract the
integral \BE\lab{bessel} \int_{0}^{+\infty} d\be
\,e^{-\rho\ch\be}=K_0(\rho)\,,\EE which is responsible for the
logarithmic behavior $\ln\frac{1}{\rho}\,$, as $\rho\to 0$. The
remaining integrals multiplying such result, \BE\lab{integrals} 2
\sum_{n=1}^{\infty}\, \frac{1}{n!}\,
\int_{-\infty}^{+\infty}\,d\be_1...d\be_{n-1}\,
\,h_n^{\al}(\be_1,...,\be_{n-1},0)\, ,\EE  give the approximate
value of the critical exponent, provided that the \lq cluster\rq
$\hspace{0.1cm}$condition \BE\lab{clcon}
h_n(\be_1,...,\be_n)=\mathcal{O}(e^{-|\be_i|})\, \EE is fulfilled,
for $\Re e\be_i\to\pm\infty$.

On the other hand, the fermionic ghost model displays a
substantial difference. The functions $h_n^\al$ depend, by
construction, on the sum of rapidities. Thus, only contributions
of even order in the series (\ref{logmua}) admit a redundant
variable, finally leading to a logarithmic behavior. The remaining
terms, of odd order, provide convergent pieces, useful to
reconstruct the normalization constant of the one-point function.

\begin{figure}[h]
\hspace{4.5cm} \psfig{figure=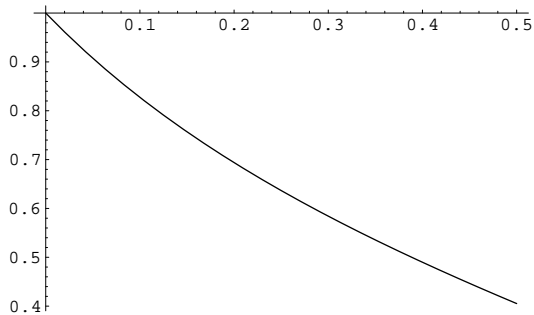,height=4.3cm,width=7cm}
\vspace{-0.25cm} \caption{-$\frac{x_\al}{\al/2}$ as a function of
the non-locality index $\al$, for
$\al\in\left[0,\frac{1}{2}\right]$. } \label{defectfig}
\end{figure}

 In order to study explicitly the
short-distance behavior of $\mu_0^\al$, we focus the attention on
the second order contribution. All we need to know is \BE\lab{h2}
h_{12}^\al(\be_1,\be_2)=-\left(\frac{\sin\pi\al}{2\pi}\right)^2\,\frac
{e^{-\left(2\al-1\right)(\be_1+\be_2)}}{\left(\ch\frac{\be_1+\be_2}{2}\right)^2}\,.\EE
Hence, substituting in (\ref{logmua}), after straightforward
calculations, we finally end up with \BE\lab{final} \ln
\mu_0^\al(t)=x_\al\ln\frac{1}{2mt}\,,\EE where the critical
exponent reads\BE\lab{expal}
x_\al=-\,\frac{1-2\al}{2\pi}\,\tan(\pi\al)\,.\EE For small values
of the non-locality index,  $x_\al\to -\al/2$. However, we are
mainly interested in the limit $\al\to1/2$, where a comparison
with the exact value $x_{1/2}\equiv x_\mu=-1/4=-0.25$, previously
derived, is possible. Eq. (\ref{expal}) leads to the result
$x_{1/2}\sim-1/\pi^2\sim-0.10$, independent of $\al$. This large
discrepancy suggests that the \lq cluster\rq
$\hspace{0.1cm}$approximation, for this particular non-locality
index, fails to reproduce the exact critical exponent with
accuracy, but, nevertheless, hints at its correct sign. Finally,
 Fig. \ref{defectfig} displays the ratio $\frac{-x_\al}{\al/2}$, in
order to make visible deviations from the small-$\al$ behavior.

\end{document}